**Harnessing electronic health records for real-world evidence**

Jue Hou, hou00123@umn.edu[1]
Rachel Zhao, rachelz@student.ubc.ca[2]
Jessica Gronsbell, j.gronsbell@utoronto.ca[3]
Brett K. Beaulieu-Jones, Brett_Beaulieu-Jones@hms.harvard.edu[4]
Griffin Webber, griffin_weber@hms.harvard.edu[4]
Thomas Jemielita, thomas.jemielita@merck.com[5]
Shuyan Wan, shuyan.wan@merck.com[5]
Chuan Hong, chuan.hong@duke.edu[6]
Yucong Lin, linyucong@bit.edu.cn[7]
Tianrun Cai, tcai1@bwh.harvard.edu[4]
Jun Wen, Jun_Wen@hms.harvard.edu[4]
Vidul A. Panickan, vidul@hms.harvard.edu[4]
Clara-Lea Bonzel, clara-lea_bonzel@hms.harvard.edu[4]
Kai-Li Liaw, kaili_liaw@merck.com[5]
Katherine P. Liao, kliao@bwh.harvard.edu[4,8]
Tianxi Cai, tcai@hsph.harvard.edu[4,9]

Affiliations
[1]Division of Biostatistics, School of Public Health, University of Minnesota, Minneapolis, MN, USA
[2]Department of Medicine, University of British Columbia, Vancouver, BC, Canada
[3]Department of Statistical Sciences, University of Toronto, Toronto, ON, Canada
[4]Department of Biomedical Informatics, Harvard Medical School, Boston, MA, USA
[5]Merck & Co., Inc., Rahway, NJ, USA
[6]Department of Biostatistics & Bioinformatics, Duke University, NC, USA
[7]Institute of Engineering Medicine, Beijing Institute of Technology, Beijing, China
[8]Division of Rheumatology, Inflammation, and Immunity, Department of Medicine, Brigham and Women's Hospital / Harvard Medical School, Boston, MA, USA
[9]Department of Biostatistics, Harvard T.H. Chan School of Public Health, Boston, MA, USA

**Correspondence:**
Tianxi Cai, ScD
Department of Biomedical Informatics
Harvard Medical School
10 Shattuck Street, Room 434
Boston, MA 02115
Ph: 617-432-4923
Fax: 617-432-0693
Email: tcai@hsph.harvard.edu

**KEYWORDS:** electronic health records, real-world evidence, data curation, medical informatics

**ABSTRACT**

While randomized controlled trials (RCTs) are the gold-standard for establishing the efficacy and safety of a medical treatment, real-world evidence (RWE) generated from real-world data (RWD) has been vital in post-approval monitoring and is being promoted for the regulatory process of experimental therapies. An emerging source of RWD is electronic health records (EHRs), which contain detailed information on patient care in both structured (e. g., diagnosis codes) and unstructured (e. g., clinical notes, images) form. Despite the granularity of the data available in EHRs, critical variables required to reliably assess the relationship between a treatment and clinical outcome can be challenging to extract. We provide an integrated data curation and modeling pipeline leveraging recent advances in natural language processing, computational phenotyping, modeling techniques with noisy data to address this fundamental challenge and accelerate the reliable use of EHRs for RWE, as well as the creation of digital twins. The proposed pipeline is highly automated for the task and includes guidance for deployment. Examples are also drawn from existing literature on EHR emulation of RCT and accompanied by our own studies with Mass General Brigham (MGB) EHR.

**INTRODUCTION**

Transforming real-world data (RWD) to real-world evidence (RWE) has the potential to augment the clinical knowledge gained from trial findings[1]. RWD offers a rich variety of clinical data over broad patient population that are often infeasible to collect in traditional randomized controlled trials (RCT). Thus, RWE generated from a larger RWD population is positioned to address questions of treatment effects across subgroups where RCTs are often underpowered, infeasible or unethical[2–5]. In contrast to RCTs which are designed to answer a specific question regarding the effectiveness of an intervention, many types RWD not structured for research, e. g. electronic health records (EHR) generated as daybook of clinical care. Thus, to effectively utilize RWD, the data curation and quality must be critically evaluated before generating RWE for regulatory purposes[6].

The Food and Drug Administration (FDA) defines RWD as data related to patient health status and/or delivery of healthcare, such as administrative claims, electronic health records (EHRs), and clinical or product registries.[7] RWE is the clinical evidence regarding the usage and benefits or risks of a medical treatment derived from RWD[7]. To accelerate the use of RWE in the "discovery, development and delivery" of medical treatments, the 21st Century Cures Act and subsequent FDA RWE framework laid the groundwork for the use of RWD in regulatory decision making, including approvals for new indications of approved drugs and post-approval requirements[7–9].

EHRs have emerged as a primary source of RWD, but present significant data quality and statistical challenges for comparative effectiveness studies[10,11]. Bartlett et al.[11] investigated the feasibility of RCT emulation with both EHRs and insurance claims and identified the lack of critical data as the major limitation. Among 220 RCTs, 85% were deemed infeasible for replication with EHR data due to the lack of readily usable structured data on (i) the inclusion and exclusion criteria, (ii) the intervention, (iii) the indication, and/or (iv) the primary endpoint. However, this evaluation was based solely on structured data (e.g. International Classification of Diseases (ICD) and Current Procedural Terminology (CPT) codes) which do not fully capture information on phenotypes, procedural interventions, indication qualifiers, imaging results, and functional disease scores required for RCTs.[11] EHRs offer a more granular view of a patient's health status in unstructured data (e. g. clinical notes, images) that is absent from structured data and can expand the availability of critical data for RWE.

RWD is also identified as foundation for the creation of digital twins[13]. The digital twins, a trending concept imported from engineering to healthcare, seek the advancement of precision healthcare by modelling and forecasting the outcomes under available interventions based on data collected from digital technologies[14] that are increasingly integrated into EHR[15]. Visionaries for digital twins advocate for automated data processing by AI (artificial intelligence), considering the complexity of future digital healthcare data and the need for real-time decision-making.

Despite the potential of EHRs in generating RWE and digital twins, the advances in medical informatics required to effectively leverage the rich information in both structured and unstructured data have not been widely adopted in the RWE community. In recent years, natural language processing (NLP) tools have been developed to extract information from various clinical notes such as signs and symptoms, lab test values, and tumor progression. AI has also been successful in medical image (e.g., CT, MRI) classification[16,17], segmentation (locating the region of interest)[18,19], and registration (merging information from multiple images)[20,21] The advancement in data curation technologies, however, still have yet to address the need for approaches to efficiently extract outcome information that cannot be accurately represented by any single EHR feature, e.g., cancer recurrence status.

Phenotyping methods that combine multiple EHR features have been developed and adopted to improve accuracy of disease status or outcome definitions, with the goal of creating a cohort of individuals with phenotypes for downstream studies. Advanced machine learning (ML) methods for phenotyping are now available for accurately and efficiently identifying patients with specific conditions, characteristics, or diseases based on the comprehensive information in their EHR as well as the temporal information of clinical events[22]. These technologies can enable reliable identification of the necessary data for RWE, but existing methods are published in technical journals outside the radar of most medical researchers. Moreover, the deployment of NLP, AI, and ML methods requires significant expertise and guidance beyond what is available in most published articles and open-source software. It is critical that deployed methods are presented in a transparent fashion where their performance can be validated, as well as with examples where the pipelines fail. Finally, subsequent analysis should incorporate robust statistical methods to minimize bias from imperfect data and confounding.

In this study, we propose to address an unmet need through emulating RCT using RWD by outlining an integrated pipeline to improve the resolution of EHR data for precision medicine research, bridging the gap between technological innovation and application to clinical studies. Incorporating state-of-art tools, we summarize the technologies and methods available for data curation and causal modelling that will enable researchers to perform robust analysis with high resolution data from EHRs for RWE generation. Our pipeline has 4 modules: 1) creating meta-data for harmonization, 2) cohort construction, 3) variable curation, and 4) validation and robust modeling (**Figure 1**). We illustrate the utility of our pipeline in the context of examples from existing literature on EHR emulation of RCT. The list of methods integrated into the pipeline are listed in **Table 1**. We showcase that the pipeline contributes simultaneously to the creation of digital twins.

**INTEGRATED DATA CURATION AND MODELING PIPELINE FOR RWE**

We begin by providing a high-level description of the related tools for each module of the pipeline. Next, we provide guidance on the deployment of the tools. Throughout this section, we frequently refer to "gold-standard labels" as the value or definition of a clinical variable curated by domain experts via manually review of the EHR of selected patients.

**Module 1: Creating Meta-Data for Harmonization**

Generating RWE relevant to a target RCT from EHR requires first curating EHR features corresponding to indication, intervention, endpoint, eligibility criteria and patient characteristics considered in the trial. Unfortunately, many clinical variables involved in RCTs are not readily available in EHRs. The first step of our pipeline, *data harmonization,* maps the clinical variables of interest to one or more relevant sources of data in EHRs. The actual *extraction of clinical variables* is described in Module 3, where potential discrepancy among multiple sources is reconciled. In existing RWE studies, exact details of how the mapping was performed are rarely reported and often cannot be easily transported to another EHR system. Domain experts might have manually mapped inclusion/exclusion criteria, e. g. if a patient was on a specific treatment, which can be labor intensive. We propose an automated procedure for data harmonization leveraging NLP to improve the efficiency and transparency of the mapping process.

We recommend the creation of meta data needed for data harmonization proceed with two key steps:

a) Concept Identification: Identify the medical concepts associated with the clinical variables from the RCT documents using existing clinical NLP software (e.g., MetaMap, NILE, cTAKES)[23–25].

The central idea of concept identification is to bring structure into unstructured free-text data by converting relevant textual information regarding clinical variables within the RCT documentation to established medical concepts through named entity recognition (NER). Medical concepts are synonym groups for clinical terms that include variations and abbreviations which capture how a clinical variable may be documented in text. The unified medical language system (UMLS)[26] maintained by the National Library of Medicine consists of over 800,000 medical concepts that are each represented by a concept unique identifier (CUI). For example, both 'rheumatoid arthritis' and the abbreviation 'RA' are mapped to the same CUI, C0003873. Different expressions for a low fever, such as 'low grade fever' and 'mild pyrexia', are mapped to the same CUI, C0239574. NER identifies medical concepts in the text extracted from EHR, such as diseases, conditions, signs and symptoms, or medications, based on the terms that are present. NER is available in many existing clinical NLP software that use concept mapping in the backend to identify the concepts that are relevant for the study[25]. The dictionary of relevant medical concepts is used as input to variable extraction in Module 3 of our pipeline.

b) Concept Matching: Match the identified medical concepts to both structured and unstructured EHR data elements.

The identified CUIs can be easily used as NLP features. For example, if the eligibility criterion includes patients with rheumatoid arthritis, then an NLP feature counting the total number of mentions of the corresponding CUI 'C0003873' can be used as a mapped NLP feature. However, since the mentions of relevant clinical variables in the unstructured text can be nonspecific, we recommend concept matching to match the identified medical concepts to associated structured EHR data, e. g. ICD codes, whenever possible. Grouping of similar structured EHR is helpful as the relationships among structured EHR variables are not reflected in existing hierarchical coding systems. Hong et al.[27] provided a standard way to group structured EHR, which produced the mapping dictionary from the group names. Specifically, four domains of codified data including diagnosis, procedures, laboratory measurements, and medications were considered. Clinical variables under any of the domains can be matched to the correspondent group by group name search. ICD codes were aggregated into PheCodes to represent more general diagnoses, e.g., MI rather than acute MI of inferolateral wall, using the ICD-to-PheCode mapping from PheWAS catalog[28]. Multiple levels of granularity of PheCode, including integer level, one-digit level and two-digit level can be utilized depending on the disease of interest. A popular alternative is the Human Phenotype Ontology (HPO)[29]. For procedure codes, including CPT-4, HCPCS, ICD-9-PCS, ICD-10-PCS (except for medication procedures), clinical classification software (CCS) categories were assigned based on the CCS mapping[30]. For medication codes, the prescription encodings at a single EHR system were aggregated to the ingredient level RxNorm codes, the normalized names for clinical drugs developed by National Library of Medicine[31]. For laboratory measurements, lab order encodings were grouped to manually annotated lab concepts or Logical Observation Identifier Names and

Codes (LOINC)[32]. The four domains of grouped structured EHR variables provide another part of the raw data for variable extraction.

It is important to note that some clinical variables, e. g. cancer stage, cancer recurrence, are poorly represented by specific structured codes and cannot be mapping to structured data. For example, cancer recurrence and cancer progression are poorly structured in EHR despite their important role in conveying a patient's status. We recommend expanding the mapping to other relevant variables using the following resources or methods. To learn the relevance of medical variables from expert curation[33,34], knowledge sources[35–39] (compiled from Wikipedia pages, journal articles, the Merck Manual, etc.) or EHR data[27,40–44], existing studies have developed 1) dictionaries of relevant variables[38,39], 2) knowledge graphs with variables as vortex and relevance as edge[33,34,45,46], in which neighboring vortexes of the target variable form the dictionary of relevant variables 3) embedding vectors with angle reflecting relevance and length reflecting frequency[35–37,40–43], from which the dictionary of relevant variables is compiled with vectors of small cosine similarities to target variable. Additionally, extracting data on clinical variables requires tools that can access raw text and image reports directly. We will describe methods to accommodate these settings in the variable extraction section (Module 3).

**Module 2: Cohort construction**

The construction of the study cohort for RWE involves identifying the patients with the condition/disease of interest, their time window for the indication and whether they underwent the interventions in the RCT. EHR data contain a large amount of data of which a subset is relevant to the study. An additional important consideration is to safeguard against risk of inadvertently including unnecessary personal health identifiers into the data for analysis. To address these 2 issues, we recommend a 3-phase cohort construction strategy that step-wisely extracts the minimally necessary data from the EHR, starting from an inclusive data mart to the disease cohort and then to the treatment arms.

Data mart

The data mart is designed to include all patients with any indication of the disease or condition of interest. To achieve the desired inclusiveness, researchers should summarize a broad list of EHR variables with high sensitivity and construct the data mart to capture patients with at least one occurrence of the listed variables. A typical choice is the disease-specific PheCode. Most PheCodes tend to be highly sensitive for the defined disease but are often not highly specific. We recommend validating the inclusiveness of the broad list by obtaining a small set of gold-standard labels reviewing for presence or absence of the phenotype sampled in a case-control manner, e. g., 20 patients selected from the data mart and 20 patients selected from outside the data mart (See Module 4). If the validation indicates that some patients are not captured by the broad list, expansion to relevant variables is available with existing resources described in Module 1. If too many patients without diseases are captured by the list, a narrower list can be considered by going one level down in the PheWAS catalog hierarchy or using specific ICD codes.

Disease cohort

After the data mart is created, the next step is to identify the disease cohort containing the subset of patients within the data mart who have the disease of interest. Identification of the disease cohort is referred to as phenotyping within the informatics literature and has been well-studied over the last decade[22,47,48]. Commonly used phenotyping tools can be roughly classified as either rule-based or machine-learning based. Rule-based approaches are simple to use, but can have poor generalization across diseases and databases, as they must be constructed in a case-by-case manner[49–51]. Machine learning approaches can be further classified as either weakly supervised, semi-supervised, or supervised based on the availability of gold-standard labels for model training. A comprehensive review of this topic was published[22]. Weakly supervised machine learning approaches have become increasingly popular as they are trained without time-consuming generation of gold standard labels. These approaches rely on so-called "silver-standard labels" that are variables that can be readily extracted for all patients in the database but are imperfect measurements for the underlying phenotype, e. g. associated PheCodes or CUIs. Examples of existing weakly supervised approaches include the anchor and learn approach[52], XPRESS[53], APHRODITE[54], PheNorm[55], MAP[56] and sureLDA[57], which all yield probabilities of the disease for each patient, rather than a deterministic classification. Semi-supervised approaches augment the silver-standard labels with a small set of gold-standard labels and are favorable when the silver-standard labels are poor measures of the underlying disease (e. g. psychological or behavior issue like suicide), e. g. AFEP[38], SAFE[39], PSST[58], likelihood approach[59], and PheCAP[60]. Supervised approaches[61,62] have given way to semi-supervised and weakly supervised methods over time, but they can be applied to new or rare diseases without established silver standard labels.

Treatment arms and timing

With a given disease cohort, one may proceed to identify patients who received the relevant treatments, which are typically medications or procedures. Most treatment information is well coded as part of structured EHR data. For example, use of linagliptin or glimepiride for type 2 diabetes (T2DM) is straightforward to identify from EHR. In addition to the specific type of treatments, the temporal order of the treatment initiation and the disease onset/progression would also play a critical role in the study. For example, the first line therapy for metastatic/recurrent cancer is defined by the pattern "metastasis/recurrence, then use of the chemotherapy before any other therapies". For such cases, it becomes necessary to ascertain both the treatment initiation time and the occurrence time of metastatic/recurrent cancer to ensure the correct temporal order. Phenotyping methods incorporating temporal order of EHR variables[63] are suitable for identification of patients matching the indication. Treatment initiation time is then typically set as time-zero in the study and then will be used for defining variables to be curated in Module 3.

**Module 3: Variable Curation**

RCT emulation with EHR data generally requires three categories of data elements: 1) the endpoints measuring the treatment effect; 2) eligibility criteria to match the RCT population; 3) confounding factors to correct for treatment by indication biases inherent in real world data. In the following, we describe the classification and extraction of the first two types while addressing the confounding in Module 4. Our classification of variables is based on three rules, format of the variable source (phenotype, text or image), its structure in EHR (well or poorly structured), as well as the need to employ phenotyping algorithms to improve its resolution. Well-structured variables have a clear mapping to structured EHR codes (e. g. diseases listed in PheWAS catalog) while poorly structured ones have none (e. g. disease progression). Even for well-structured data elements, there may be a need to improve the accuracy of a clinical variable, such as the disease status as discussed in Module 2, due to noisiness of EHR codes. We group eligibility criteria and confounding factors together, as they are covered by the general pre-treatment baseline variables.

<u>Baseline eligibility criteria</u>

The list of eligibility criteria is provided by the RCT protocol and mapped to corresponding EHR variables in Module 1. If a user would like to perform population adjustment (e.g., weighting or matching), the list of variables available in RCT data or reported in the corresponding paper can be used.

We classify baseline variables into 3 types: *phenotype-derived*, *text-derived,* and *image-derived*. Extraction of *phenotype-derived* variable is essentially employing a phenotyping algorithm as discussed above. If a variable is well structured, we may use its EHR indicators as silver standard labels in unsupervised or semi-supervised phenotyping methods. Otherwise, only supervised methods can be applied.

The extraction of the other two types may require specific tools. *Text-derived* variables include numerical data embedded in clinical notes with a tag such as a relevant concept or code in the vicinity. The tool EXTEND was developed to link the numbers to their tags, which has been applied for body mass index, ejection fraction, vital signs, and performance status (Eastern Cooperative Oncology Group or Karnofsky Performance Scale) with high accuracy[64]. A context sensitive variant (NICE) was developed to disambiguate common features like stage for the disease of interest. NICE can also extract radiological or genetic information, e. g. tumor size and mutation variant, from text reports along with a relevant date if the note pointed to a past event[65]. RCTs tend to adopt rigorous radiological evaluation criteria in the eligibility, e. g. diameter of cancer tumor in Response Evaluation Criteria in Solid Tumors (RECIST)[66]. However, we discovered that such evaluation was rarely measured and documented in real-world radiological reports, as reported by other study[67]. With the advancement of image recognition technology, extraction of *image-derived* evaluation from imaging data in EHR becomes possible. Segmentation tools have been developed for organs[68], blood vessel[69], neural system[70,71], which may produce the physical measures. Diagnostic tools have been developed for nodule detection[72,73], cancer staging[74], and fractional flow reserve[75,76].

A preliminary emulation cohort can be constructed from the extracted eligibility criteria. Users may use a relaxed or conservative rule depending on the anticipated sample size. Later in Modules 3 and 4, further modification will be applied to finalize the emulation.

Endpoints

The extraction of endpoints varies across types of endpoints. We classify the endpoints into 4 types: *death*, *binary*, *time-to-event* and *numerical*. Death is singled out for its external source.

*Death* information can be obtained by linking EHR to national vital statistic databases. Caution should be taken on possible data leakage or informative censoring, even for presumably reliable endpoints like death. We noticed missing death status from terminal stage patients, likely due to out of state home hospice. In that case, the endpoint should be modified to as in-hospital death or discharge in a terminal condition. Discharge in terminal conditions can be extracted as typical binary phenotypes by semi-supervised methods using EHR data of the last month before loss-to-follow-up.

*Binary* endpoints are essentially a binary status of presence or absence of a clinical condition during or at the end of the follow-ups, e. g. 1-year remission of the disease. Thus, they can be extracted by phenotyping methods using the EHR data since treatment initiation. As many endpoints are disease progressions rather than diagnoses, they are poorly structured. Consequently, semi-supervised phenotyping methods aggregating auxiliary information from other relevant features may be preferred to balance between resource needed to manually curate gold standard labels via chart review and the accuracy of the final endpoint definition.

*Time-to-event* endpoints include many common primary endpoints, e. g. progression free survival for cancer. The longitudinal trajectories of EHR features (e. g. diagnosis and procedures) relevant to the event of interest provides information on the event time through incidence phenotyping. Incidence phenotyping also includes various unsupervised[77,78], semi-supervised[79,80], supervised[81,82] approaches.

*Numerical* endpoints, including both ordinal endpoints like disease severity scores and real number endpoints like tumor size, are usually difficult to extract from EHR. Tools for *Text-derived* baseline variables are options for extraction, but missing documentation in the real-world setting imposes the intrinsic difficulty. If measurement is not captured at the specific timing of interest, some temporal tolerance should be considered. Effort has been put into data-driven construct of severity scores from EHR for depression, multiple sclerosis, stroke, in which machine learning trained the EHR severity score on a labeled subset with standard severity scores from registry, questionnaire or NLP tool.

Missing Data

Missing data is a common issue for RWD. Certain information could be absent in real-world medical records, and thus not available even from manual abstraction. For diagnosis baseline variables and endpoints, the absence of records on the diagnosis of interest usually suggests the diagnosis never observed for the patient, so it may be counted as negative for the diagnosis. For *text/image-derived* baseline variables, *numerical* endpoints or lab testing results, the absence of extraction should be marked as missing data. In the downstream analysis, standard strategies can be employed to handle the missing data, imputation, or missing-indicator. Caution should be given to informative missing. If the missing rate is too high, compromise must be considered for the missing variables like discarding from baseline variables or finding surrogates for endpoints. Sensitivity analyses can be performed to ensure that the results are consistent across different strategies for handling missingness.

**Module 4: Validation and Robust Modelling**

Inaccurate data curation and confounding can lead to biased RWE. Even with reasonably accurate medical informatics tools, remaining errors from data curation will be carried over to downstream analyses, potentially causing bias in treatment assessment. Confounding is a constant challenge for assessing treatment with observational data[83], including the routinely collected EHR. Confounding factors, variables that affect both the treatment assignment and outcome, must be properly adjusted. To minimize the bias, the pipeline should include 1) validation for optimizing the medical informatics tools in Modules 2 and 3 ; 2) analyses robust to remaining data error[84–86]; 3) comprehensive confounding adjustment[87–89].

Validation and Tuning of Data Curation

First, we suggest validating the quality of data curation by detecting inconsistency between annotation and extraction. When validation of all variables is infeasible, priority should be given to variables defining: 1) the indication and eligibility; 2) treatment arms; 3) end points; 4) key confounding variables. To ensure sufficient detection chance, we recommend the validation sample size formula

$$Validation\ Size \geq \log(1 - Detection\ Chance)/\log(1 - Error\ Tolerance).$$

Users can choose the detection chance and error tolerance according to the context and report with analysis. With 95% detection chance and 5% error tolerance, a subset of at least 59 is needed. The validate set can be used for tuning the data curation when excessive error is detected. To avoid overfitting, we recommend use two validate sets, one for tuning another for post-tuning re-validation.

Robust analysis for imperfect data

Second, three annotations should be created for emulation cohort robust downstream analysis:

1. Indicator for indication, arm, and eligibility. Besides the levels for the treatments of interest, a level for exclusion should be created for ineligible patients.
2. Actual end points consistent with any modification as in Module 3.

3. Other variables for population adjustments.

The size of this subset should be determined by the recommend sample size of the supervised or semi-supervised methods in the downstream analysis. The annotations for variables with error in validation are created for this larger set. We describe a sampling scheme that efficiently recycles the annotation in Supplementary Material **eSection 2**.

Robust adjustment for confounding

The list of confounding factors, however, is seldom known a priori. A common strategy in RWD treatment effect analysis is to include many probable confounding factors and capture the confounding with model selection techniques[87,88]. Here, we provide a broad list for creating a list of potential confounding factors:

- *Demographic* data of the RCT is routinely described in the RCT reporting paper. A list can be pulled from there.
- Some *eligibility criteria* defining variables may have multiple eligible levels or values. They usually carry clinical importance, and thus likely to affect both treatment and outcome in real-world practice.
- General *medical history* is described by the disease and symptom diagnoses, which covers the comorbidities. The diagnosis codes at baseline grouped into integer level PheCodes can be used.
- *Disease history* covers the disease severity, course of progression and past treatments. Both an expert defined approach and data-driven approach can be considered. The expert defined list may come from a domain expert or existing literature on related observational studies. The data-driven list is generated like the mapping to relevant variables in Module 1.

- *Risk factors*, variables affecting outcomes, contain all confounders. A review of literature on the disease will provide a list of identified risk factors.
- *Calendar year of treatment initiation*. If the treatment initiation times in EHR cover a long time span or landmark change in practice, the calendar year may become the confounding factor[88].

Validation is not applicable for the large number of candidate confounding EHR factors because they are defined within EHR and thus exact. In the downstream analysis, we recommend the doubly robust estimation that produces accurate treatment assessment if either mechanism of treatment assignment (propensity scores) or outcome (outcome regression) is properly modelled[90].

**Creation of Digital Twins from RWD**

The notation of digital twins has strong resemblance with established concepts in causal inference, e. g. potential outcomes[91] and virtual twins[92]. In essence, the depiction of digital twins in precision healthcare setting is the personalized optimization over available interventions according to their forecasted outcomes derived from modelling the outcome mechanism[14], which would be a byproduct of the robust causal modelling. Consequently, the RWD created by our data curation pipeline can also be used to form the digital twins that supplement the existing precision medicine studies conducted over RCT data[93]. For clinical development in general, this can further improve understanding of treatment heterogeneity and inform study design.

**Data Availability**

Data sharing not applicable to this article as no datasets were generated or analysed during the current study.

**DISCUSSION**

The described data curation and modeling pipeline has broad potential for the application of RWD in clinical development, for example using RWD as the external/hybrid control arms or conducting pragmatic trials with RWD. For the former, the external control could be used as a benchmark for a single arm design, or instead be used to augment an existing RCT control arm which allows for better control of potential bias. Specifically, the described pipeline could (1) better identify patients that match the target trial (along with an assessment of any discrepancy), (2) encourage harmonization between RCT and RWD variables (which aids in statistical adjustments), (3) address missing data issues (which are more prevalent in RWD) through efficient imputation strategies, and (4) extract more relevant variables by leveraging both structured and unstructured data. Overall, the pipeline aims to develop a fit-for-purpose RWD dataset through a robust and transparent process. The pipeline may also be used to generate RWD for other purposes. For example, existing observational study cohorts can be expanded or refreshed by the RWD generated from EHR, which would enhance the usability of RWD on other fronts like safety. While there may be areas for potential improvements and in some cases the RWD may not be appropriate for the question/study of interest, the described pipeline lays a roadmap to format RWD geared toward generating RWE enabling downstream applications that could potentially accelerate clinical development, and ultimately, improve patient care.

**Conditions Needed to Implement**

To deploy the pipeline, the following conditions are required. First, the EHR infrastructure should allow mapping of local codes to common ontologies for the structured data (e.g., ICD, CPT, RxNorm). Second, the available medical notes (and images if necessary) must provide sufficient information for medical experts in the research team to ascertain the clinical variable of interests. This ensures that the narrative notes along with other codified data capture most routinely collected clinical information. However, variables specifically intended for clinical trials may not be available, e. g. performance status. If patients receive care at multiple healthcare centers or non-prescription medications, data from a single institution may not accurately capture the relevant clinical information due to well-known data leakage issue[5,94]. The discrete nature of medical encounters might also limit the precision of temporal information about events occurring between visits. Third, the automated extraction tools have reasonable performance for the clinical variables of interest. Otherwise, no additional information can be gained outside the annotated subset.

**Limitations**

For clinical development applications (ex: external control arms), consistent with guidance for traditional RCTs, RWD related statistical analysis plans need to be pre-specified and should be discussed with relevant regulatory agencies. Similarly, the data curation plans should be prespecified to the extent possible. It is important to ensure the reliability (data accuracy, completeness, provenance, and traceability) and relevance (key data elements, ex: treatment line) used to help support a regulatory decision. The proposed data curation directly tackles this requirement, as any phenotyping or variable/outcome extraction can employ an automated process under the proposed framework. However, even with an automated data curation process, there are potential limitations. First, pre-approval treatments can be nonexistent or scarce in EHR. If one arm of the RCT is placebo or futile experimental treatment, response data following the treatment may not be available in RWD. In addition, it is generally difficult if not infeasible to emulate RCT comparing an effective novel therapy to clearly inferior treatments since the treatment-by-indication bias can be extreme. Second, medication dose and regimen administration patterns are usually not documented in EHR. RCTs comparing doses or administration patterns of the same medication may not be emulated. Third, it may not be possible to extract the clinical outcome of interest. For example, in oncology studies, it may not be feasible to extract tumor response based on RECIST 1.1 due to lack of routine documentation and it may be reasonable to consider an alternative, but still validated, response metric. Fourth, infrequent

RWD encounters may limit the precision of time-to-event endpoints in RWD. Event reported at a later RWD encounter may have developed at any time since the previous RWD encounter. If the encounter frequency in RWD is much lower than monitor frequency of the target RCT, analysis for interval censoring data should be considered. Fifth, transferring across health care systems may cause incomplete treatment or endpoint information within each health care system. Such data leakage issue can be overcome by data consortium across health care systems equipped with privacy-preserving and communication-efficient federated learning tools. Lastly, imperfect extraction of key variables, such as confounding variables and clinical inclusion/exclusion criteria, may induce population or confounding bias. Domain expert knowledge can greatly enhance this step by identifying the key variables of interest for a given target study and guiding validation. We recommend combining expert knowledge along with automated processes. Adjusting for high-dimensional baseline covariates with proper variable selection may alleviate the confounding bias due to imprecise confounder curation.

**ACKNOWLEDGEMENT**

**Author Contributions**

J. H. led the design of the protocol, performed the construction of modules, figure/table generation, and writing. R. Z., B. K. B.-J., G. W., T. J. performed reviews on the background of the study and drafted the introduction and discussion accordingly. C. H., Y. L., Tianrun C., J. W., V. A. P. performed the reviews on informatic tools in Modules 1-3 and drafted their descriptions therein. J. G. and C.-L. B. performed proof reading, content curation and figure/table generation. K.-L. L., K. P. L. and Tianxi C. conceived the protocol, oversaw the review process, and provided necessary feedback, proof reading, and content curation.

**Competing Interests**

The Authors declare no Competing Non-Financial Interests but the following Competing Financial Interests. The study is funded by Merck & Co., Inc., Rahway, NJ, USA. Katherine P. Liao is funded by National Institutes of Health P30 AR072577. Brett Beaulieu-Jones is funded by National Institutes of Health K99NS114850.

**Role of Funder**

The funding agency (Merck & Co., Inc., Rahway, NJ, USA) of the work contributed to the design and conduct of the study; preparation, review, or approval of the manuscript; and decision to submit the manuscript for publication. Merck & Co., Inc., Rahway, NJ, USA played no role in the collection, management, analysis, and interpretation of the data. National Institutes of Health played no role in design and conduct of the study; preparation, review, or approval of the manuscript; and decision to submit the manuscript for publication; collection, management, analysis, and interpretation of the data.

**FIGURE CAPTIONS**

**Figure 1. The Integrated Data Curation pipeline** designed to enable researchers to extract high quality data from electronic health records (EHRs) for RWE.

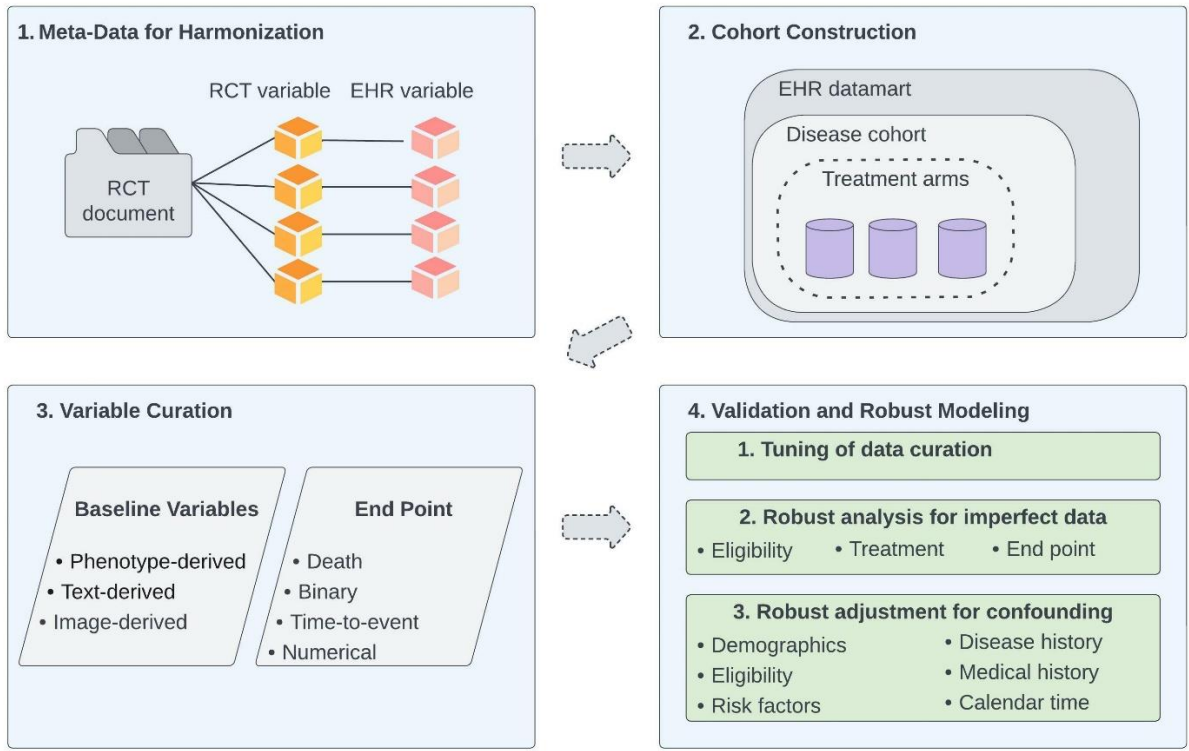

**Table 1. Summary of methods in each step by their use.**

| Module | Step | Use | Methods |
|---|---|---|---|
| **Data harmonization** | Concept Identification | *Identify medical concepts from RCT documents* | Metamap[23]/HPO[29], NILE[24], cTAKES[25], |
| | Concept Matching | *Grouping of structured EHR* | PheWAS catalog[28], CCS[30], RxNorm[31], LOINC[32] |
| | | *Expansion and selection of relevant features using knowledge source or cooccurrence* | Expert curation[33,34], knowledge sources[35–39], EHR data[27,40–44] |
| **Cohort Construction** | Data Mart | *Filter patients with diagnosis codes relevant to disease of interest* | PheWAS catalog[28] or HPO[29] |
| | Disease Cohort | *Identify patients with disease of interest through phenotyping* | Unsupervised: anchor and learn[52], XPRESS[53], APHRODITE[54], PheNorm[55], MAP[56] and sureLDA[57]; Semi-supervised: AFEP[38], SAFE[39], PSST[58], likelihood approach[59], and PheCAP[60] |
| | Indication and Treatment Arm | *Identify indication conditions before treatment* | Phenotyping with temporal input[63] |
| **Variable Extraction** | Extraction of Baseline Variables or Endpoints | *Extraction of binary variables through phenotyping* | Same as *Identify patients with disease of interest through phenotyping* |
| | | *Extraction of numerical variables through NLP* | EXTEND[64], NICE[65] |
| | Extraction of Baseline Variables | *Extraction of radiological characteristics through medical AI* | For organs[68], blood vessel[69], neural system [70,71], nodule detection [72,73], cancer staging[74], fractional flow reserve[75,76] |
| | Extraction of Baseline Endpoints | *Extraction of event time through incidence phenotyping* | Unsupervised [77,78], Semi-supervised [79,80], Supervised[81,82] |
| **Downstream Analysis** | | *Efficient and robust estimation of treatment effect with partially annotated noisy data* | SMMAL[86] |

# Supplementary Materials

**eSection 1.** Workflow and example for Module 2.

As the Module 2 (cohort construction) is generally similar across studies, we provide a detailed workflow with a use case on colorectal cancer study[1].

1. Build the data mart. Map the disease to its associated PheCode. Extract the patients with the diagnosis code under the PheCode from database. For example, to create a data mart for colorectal cancer, all subjects with PheCode 153 for colorectal cancer would be included.

2. Construct the disease cohort. Many phenotyping algorithms require the silver-standard labels, often the total counts of associated PheCodes or CUIs. Some also require a feature that is a proxy for healthcare utilization to account for the heterogeneity in the dataset. A set of gold-standard labels should be generated for validating the performance of the phenotyping algorithm. Additional gold-standard labels are needed for training supervised or semi-supervised phenotyping methods. Our example is based on MAP[1,2].

    a. Extract the silver-standard labels and healthcare utilization feature for patients in data mart. We recommend using total days with disease PheCode and total number of disease CUI as silver-standard labels and total days with any ICD code as healthcare utilization. Our recommendation stems from the observation that multiple codes in one day merely reflect the administration pattern (less for integrated provider and more for segmented providers) yet multiple mentions of the disease in medical notes usually indicate likelihood of disease onset. Manual chart review to obtain the gold-standard labels for a random subset, e. g. 59 patients as in Module 4, should also be done in parallel.

    b. Apply an unsupervised phenotyping method, e. g., MAP, and validate the performance with the gold-standard labels. If the numeric prediction is reasonable (area-under-receiver-operating-characteristic-curve, AUROC >0.9),

choose the threshold with 0.95 specificity and construct the disease cohort with patients whose numeric prediction is greater than the threshold. Otherwise, go to next step.

c. Extract EHR variables from the expanded mapping and generate additional gold-standard labels, e.g., 200 patients. Run a semi-supervised phenotyping method, e.g., PheCAP.

For phenotyping of colorectal cancer patients, the silver standard labels can be the diagnosis codes under PheCode 153 and the main "colorectal cancer" CUI C0009402. The healthcare utilization can be measured by the number of days with ICD codes.

3. <u>Create the treatment arms.</u> Most medication or procedure based treatments are mapped to structured EHR codes. With the mapping established in Module 1, we can create treatment arms of patients with the correspondent treatment code. For example, patients with colectomy or laparoscopy-assisted colectomy procedure codes formed the open colectomy arm and the laparoscopy-assisted colectomy arm[1]. The creation of treatment arms may involve ascertaining the temporal order of disease onset/progression and the treatment, like the "first line therapy for metastatic/recurrent cancer" example mentioned in Module 2. For chemotherapies, only those administered after metastasis/recurrence qualified for the indication, excluding the adjuvant chemotherapies. For such studies, the temporal phenotyping[3] workflow includes the following steps:

   a. Extract the encounter level EHR variables informative for treatments of interest and the disease onset/progression in indication. For each patient, create the marginal counts for each variable and the ordered counts for each pair (SPM and tSPM)[3].

   b. Create the silver standard label to perform features screening and model selection. Order counts for treatment of interest variables after indication disease onset/progression variables are natural choices for the silver standard label. Other rule-based extraction can also be used here.

c. Train the temporal phenotyping model with selected features on gold-standard labels of expert annotated treatment and indication. The numeric probabilities are produced.

d. Determine the optimal threshold for numeric probability over another set of gold-standard labels with minimal miss-classification.

To identify first line therapy for metastatic colorectal cancer, some gold standard labels on the first line use of the targeted therapy or chemotherapy of interest. Silver standard labels may be constructed from ordered counts of any indicators for metastasis (NICE[4] extraction of metastasis or cancer staging, PheCode 198) before the indicators for therapies (NLP extraction of regimen and medications, medication codes). Marginal counts and ordered counts for variables mapped to "metastasis" and the list of targeted therapies or chemotherapies of interest.

4. Ascertain treatment initiation time. The time-zero is obtained as the date of first treatment indicator or the first *qualified* treatment indicator if temporal phenotyping is involved. For the colectomy example, the time-zero is the date of first colectomy or laparoscopy assisted colectomy code. For the first line therapy for metastatic cancer, the time-zero is the date of first code of targeted therapies or chemotherapies after indicator for metastasis.

**eSection 2.** Efficient Sampling scheme for Module 4.

As annotations are needed at various places, from cohort construction through variable extraction to validation, a strategy to reuse them may maximize the utility of the time and labor put into annotation. To facilitate robust analysis, it is important to annotate all error-prone variables over the same subset for downstream analysis. Aiming for this goal, we propose the following sampling scheme.

1. Randomly shuffle the initial data mart. Whenever k annotations are needed during the cohort construction and variable extraction, always choose the first k qualified patients according to the post shuffling order.
2. Prioritize the extraction of eligibility criteria so that the extraction of other variables may be focused within the preliminary emulation cohort.

   The scheme maximizes the overlapping of the annotated sets for various tasks so that all annotations qualified at a later stage can be reused.

**Reference.**